\documentclass[preprint2]{aastex}

\shorttitle{A log-quadratic $M_{\rm bh}$--$n$ relation}
\shortauthors{Graham \& Driver}

\begin{document}

\title{A log-quadratic relation for predicting supermassive black 
hole masses from the host bulge S\'ersic index}

\author{Alister W.\ Graham\altaffilmark{1}}
\affil{Mount Stromlo and Siding Spring Observatories, Australian 
National University, Private Bag, Weston Creek PO, ACT 2611, Australia.}
\altaffiltext{1}{Corresponding Author: Graham@mso.anu.edu.au}

\author{Simon P.\ Driver\altaffilmark{2}}
\affil{School of Physics \& Astronomy, University of St Andrews, 
North Haugh, St Andrews, Fife, KY16 9SS, UK}
\altaffiltext{2}{Scottish Universities Physics Alliance (SUPA).}

\begin{abstract}

We reinvestigate the correlation between black hole mass and bulge
concentration.  With an increased galaxy sample (totalling 27), 
updated estimates of 
galaxy distances, black hole masses, and S\'ersic indices $n$ --- a
measure of concentration --- we perform a least-squares regression
analysis to obtain a relation suitable for the purpose of predicting
black hole masses in other galaxies.  In addition to the linear
relation, $\log (M_{\rm bh}) = 7.81(\pm 0.08) + 2.69(\pm
0.28)\log(n/3)$ with $\epsilon_{\rm intrinsic}=0.31^{+0.09}_{-0.07}$
dex, we investigated the possibility of a higher order $M_{\rm bh}$--$n$
relation, finding the second order term in the best-fitting quadratic 
relation to be inconsistent with a value of zero at greater than the 99.99\%
confidence level.  The optimal relation is given by $\log (M_{\rm bh})
= 7.98(\pm 0.09) + 3.70(\pm 0.46)[\log(n/3)] - 3.10(\pm
0.84)[\log(n/3)]^2$, with $\epsilon_{\rm intrinsic} =
0.18^{+0.07}_{-0.06}$ dex and a total absolute scatter of 0.31 dex.
Extrapolating the quadratic relation, it predicts black holes with
masses of $\sim 10^3 M_{\sun}$ in $n=0.5$ dwarf elliptical galaxies,
compared to $\sim 10^5 M_{\sun}$ from the linear relation, and an
upper bound on the largest black hole masses in the local universe,
equal to $1.2^{+2.6}_{-0.4} \times 10^9 M_{\sun}$.

In addition, we show that the nuclear star clusters at the centers of
low-luminosity elliptical galaxies follow an extrapolation of the same
quadratic relation --- strengthening suggestions for 
a possible evolutionary link between 
supermassive black holes and nuclear star clusters.  Moreover, we
speculate that the merger of two such nucleated galaxies, accompanied
by the merger and runaway collision of their central star clusters,
may result in the late-time formation of some supermassive black
holes.  
%

Finally, we predict the existence of, and provide equations for, 
an $M_{\rm bh}$--$\mu_0$ relation, in which $\mu_0$ is the 
(extrapolated) central surface brightness of a bulge. 

\end{abstract}


\section{Introduction}

While perhaps not surprising to AGN-astronomers, over the past five to ten 
years, the notion that a supermassive black hole (SMBH; $M_{\rm bh} \ge 10^6
M_{\sun}$) resides at the heart of every significantly large
($M_{\rm bulge} \ge 10^9 M_{\sun}$) galaxy bulge, even inactive bulges, has
changed from a dubious idea to a mainstream belief. 
Support for the tide-of-opinion change has, in part, arisen from studies of
our own galaxy. 
Sch\"odel (2003) and Ghez et al.\ (2005, and references therein, see
also Broderick \& Narayan 2006) have shown that the mass required
inside of the innermost resolved volume of the Milky Way is
sufficiently large that it rules out alternatives to a SMBH, such as a
cluster of dead stars or stellar mass black holes which, if they did
once exist, have surely now merged to form a single, massive object
(e.g., Miller 2006).  Although there is only one other sufficiently
well resolved galaxy where such a conclusion can be drawn, NGC~4258
(Miyoshi et al.\ 1995),
it seems reasonable to accept that the dark concentrations of mass at
the centers of other galaxies are also SMBHs, and we adopt this
convention, or at least terminology, here.

Roughly some three dozen galaxies are close enough that their central,
SMBH mass has been measured directly through its influence on the
motion of the surrounding gas and stars (Kormendy \& Gebhardt 2001;
Merritt \& Ferrarese 2001a; Ferrarese \& Ford 2005, their
Table~II)\footnote{We do not use the eight galaxies listed in part two
of Table~II from Ferrarese \& Ford 2005 for which the SMBH masses
might be in error.}.  To acquire the mass of the central dark object
in other (inactive) galaxies requires an alternative approach, and a
number of indirect means to do so have been proposed (e.g., Ferrarese
\& Merritt 2000; Gebhardt et al.\ 2000; Marconi \& Hunt 2003; Novak,
Faber \& Dekel 2006).

Perhaps the most often used relation --- due to the small level of scatter
($\sim$0.3 dex) --- is the $M_{\rm bh}$--$\sigma$ relation (Ferrarese
\& Merritt 2000; Gebhardt et al.\ 2000).  With an equivalent level of
scatter, and the advantage of requiring only galaxy images rather than
spectra, is the $M_{\rm bh}$--$n$ relation (Graham et al.\ 2001,
2003a), where $n$ is a measure of the concentration of the stars within the 
bulge.  More precisely, $n$ is the inverse exponent from the
best-fitting S\'ersic (1963) $R^{1/n}$ light-profile (see Graham
\& Driver 2005 for a review of this model). 
Other relations, often reported to have more scatter than the above
two relations, are the $M_{\rm bh}$--$L$ and the $M_{\rm bh}$--$M_{\rm bulge}$
relations (Kormendy 1993; Magorrian et al.\ 1998).  Although, recent studies 
which have excluded the disk-dominated spiral galaxies (e.g., McLure
\& Dunlop 2002), or used only the bulge luminosity from the disk galaxies
after performing an accurate bulge/disk decomposition (e.g., Erwin et
al.\ 2002; Marconi \& Hunt 2003) have obtained a relationship with a
similarly low level of scatter (see also H\"aring \& Rix 2004). 
All of these empirical relations can be, and indeed must be,
used to constrain any {\it complete} theory or model of galaxy/SMBH
co-evolution.  Furthermore, all of these relations can be used
to gauge the mass of SMBHs in other galaxies.

In this paper we use the S\'ersic index $n$, together with updated SMBH
masses (Section~2), to derive the first $M_{\rm bh}$--$n$ relation
constructed for the purpose of predicting $M_{\rm bh}$ in other
galaxies (Section~3).   This relation may differ from the
intrinsic, astrophysical relation presented in Graham et al.\ (2003a)
due to the different method of regression that is required. 
Armed with such a relation, one requires only images --- which need
not even be photometrically calibrated --- to predict accurate SMBHs
in other galaxies.  Additional advantages with the use of a global
measure such as $n$, are that it is not heavily affected by possible
kinematical sub-structure at the center of a bulge, nor by rotational
velocity or vertical dispersion of an underlying disk, nor aperture
corections.  Furthermore, the quantity $n$ is cheap to acquire in
terms of telescope time and unlike absolute magnitudes and masses, it 
does not depend on galaxy distance nor an uncertain mass-to-light ratio.

As noted in Graham et al.\ (2001), there is no {\it a priori} reason
to presume that the relation between $\log M_{\rm bh}$ and $\log n$ is
linear, and we therefore explore the suitability of a 
a quadratic 
equation.  In so doing, we find the second order term is inconsistent
with a value of zero at the 99.99\% confidence level
(Section~3.2).  Implications at the low- and high-$n$ end of the
relation are discussed, relative to the optimal linear fit.

In a forthcoming paper we will apply this quadratic relation to the
Millennium Galaxy Catalog (Liske et al.\ 2003; Cross et al.\ 2004;
Driver et al.\ 2006) containing 10 095 galaxies --- which have been modeled
with a S\'ersic-bulge plus exponential-disk (Allen et al.\ 2006) ---
to determine the local supermassive black hole mass function 
(e.g., Salucci et al.\ 1999; Yu \& Tremaine 2002; Granato et al.\ 2004; 
Shankar et al.\ 2004) 
and space density in both early- and late-type galaxies.  

In Section~4 we expand the $M_{\rm bh}$--$n$ diagram into an $M_{\rm
mco}$--$n$ diagram, where $M_{\rm mco}$ is the mass of the central
compact object, which may be a SMBH or a nuclear star cluster.  We
show that the nuclear star clusters in early-type galaxies appear to
follow the curved $M_{\rm bh}$--$n$ relation defined by galaxies with
SMBHs, and we discuss some of the implications this may entail.  In
Section~5 we show that our $M_{\rm bh}$--$n$ relation is independent
of the Hubble constant.  

Given there are now several relations between SMBH mass and the
properties the host bulge, in Section~6 we briefly present some
musings as to what may be the fundamental primary relation.  In this
section we derive a new set of equations relating the SMBH mass to the
(extrapolated) central surface density of the host bulge.  Finally, in
Section~7 we provide a brief summary of the paper.

\section{Data for the $M_{\rm bh}$--$n$ relation}


Our sample is comprised of the 21 galaxies used in Graham et
al.\ (2001) plus an additional six new galaxies.  

A discussion of the
original galaxy light-profiles can be found in Erwin et al.\ (2002),
see also Trujillo et al.\ (2004).
Due to updated galaxy distances and SMBH mass measurements, and a
refined analysis of the major-axis
light-profiles, some of the 27 data points (14 E and 13 S0/Sp) given
in Table~\ref{Table1} are slightly different to those published in
Graham et al.\ (2001, their table 1) and shown in Graham et al.\
(2003a, their figure 1).  However, only for two galaxies has the S\'ersic index
changed by more than 20\% --- the typical uncertainty on this index
(e.g., Caon, Capaccioli, \& D'Onofrio 1993). 
We briefly comment on these two galaxies here and present their new
S\'ersic fits in Appendix~A.

Although the bulge of the Milky Way is nowadays recognised as having an
exponential ($n=1$) light-profile, we have fitted the near-infrared
data from Kent, Dame, \& Fazio (1991) and report an index of $n=1.32$
rather than exactly 1.  The second galaxy in question is NGC~4564.  Subsequent to
the analysis in Trujillo et al.\ (2004), NGC~4564 is now recognized as
an S0 galaxy that was mis-classified as an E galaxy.  An
$R^{1/n}$-bulge plus exponential-disk decomposition has therefore been
performed.  This resulted in its bulge S\'ersic index $n$ increasing
from 2.1 to 3.2.

Four of the six new galaxies are listed in Ferrarese \& Ford (2005)
(NGC~3115, 4486, 4649, 4697) and our S\'ersic fits to their
light-profiles are also presented in Appendix~A.  In addition, we have used
the bulge/disk decomposition shown in Graham (2002) for the
supposedly ``compact Elliptical'' galaxy NGC~221 (M32), which was not
included in our original paper.  We have also been able to include
NGC~1399 (Houghton et al.\ 2006), which previously had no direct SMBH
mass estimate.  This galaxy's light-profile has been modeled in
D'Onofrio, Capaccioli, \& Caon (1994) and we adopt their S\'ersic index
having obtained the same value from our own fitting of their data.

Despite our attempt to include all of the lowest and highest mass
SMBHs, as these data points can have substantial weight on any fitted
relation, we have not included the peculiar elliptical galaxy IC~1459
due to uncertainty on both its SMBH mass and its S\'ersic index $n$.
This galaxy displays clear signs of past interaction, evidenced by
stellar tidal tails (Malin 1985) and stellar shells and ripples at
large radii (Forbes \& Reitzel 1995). Due to its unrelaxed, disturbed
morphology, its light-profile is not well fitted with an $R^{1/n}$
model.  It also possesses a fast counter-rotating stellar core.  While
the stellar dynamics of the core suggest a SMBH mass of $2.6 \times
10^9 M_{\sun}$, the gas dynamics reveal the mass could be as low as
$3.5 \times 10^8 M_{\sun}$ (Cappellari et al.\ 2002).

The upper section of Table~II in Ferrarese \& Ford (2005) lists 25
galaxies with SMBH masses derived from resolved dynamical studies.  A
further five galaxies in this list have marginally resolved mass
estimates, with $0.39 < r_{\rm sphere~of~influence}/r_{\rm resolved}<
0.92$.  Aside from IC~1459, we are only missing quality $R$-band
images for three of these 30 galaxies: NGC~3608, NGC~5128
(Centaurus~A) and the relatively distant ($z=0.056$) galaxy Cygnus~A.

Most of our 27 galaxies have distances from Tonry et al.\ (2001), and 
updated SMBH mass estimates have come from Tremaine et al.\ (2002), 
with exceptions noted in Table~\ref{Table1}.  
Three galaxies in our sample do not have `Surface Brightness
Fluctuation' distance 
measurements in Tonry et al.\ (2001).  For the relatively
distant galaxies NGC~6251 and NGC~7052, we used their heliocentric
velocities (Wegner et al.\ 2003)
and a Hubble constant $H_0 = 73$ km s$^{-1}$ Mpc$^{-1}$ (Blakeslee et
al.\ 2002) --- consistent with the HST Key project value of 72$\pm 3 \pm
7$ (Freedman et al.\ 2001) and the WMAP value of $73^{+3}_{-3}$
(Spergel et al.\ 2006).  For the remaining Virgo cluster galaxy
NGC~4342, we used the mean Virgo cluster distance of 17.0 Mpc (Jerjen,
Binggeli, \& Barazza 2004).  The above three galaxies have slightly
different SMBH masses in Table~1 to those in Tremaine et al.\ (2002)
due to the slightly different distances used.


\begin{deluxetable}{lccc}
\tablewidth{530pt}
\tablecaption{Galaxy Parameters\label{Table1}}
\tablehead
{
\colhead{Galaxy} & \colhead{Distance} & \colhead{$M_{\rm bh}$}      & \colhead{$n$} \\
\colhead{}       & \colhead{Mpc}      & \colhead{($10^8 M_{\sun}$)} & \colhead{}
}
\startdata
\multicolumn{4}{c}{Elliptical Galaxies} \\
NGC  821       & 24.1  &  0.85$^{+0.35}_{-0.35}$  &  4.00$^{+0.80}_{-0.67}$  \\
NGC 1399$^{a}$ & 20.0  &  12$^{+5}_{-6}$     &  16.8$^{+3.36}_{-2.80}$   \\
NGC 3377       & 11.2  &  1.00$^{+0.9}_{-0.1}$    &  3.04$^{+0.61}_{-0.51}$  \\
NGC 3379       & 10.6  &  1.35$^{+0.73}_{-0.73}$  &  4.29$^{+0.86}_{-0.72}$  \\
NGC 4261       & 31.6  &  5.20$^{+1.0}_{-1.1}$    &  7.30$^{+1.46}_{-1.22}$  \\
NGC 4291       & 26.2  &  3.10$^{+0.8}_{-2.3}$    &  4.02$^{+0.80}_{-0.67}$  \\
NGC 4374       & 18.4  &  4.64$^{+3.46}_{-1.83}$  &  5.60$^{+1.12}_{-0.93}$  \\
NGC 4473       & 15.7  &  1.10$^{+0.40}_{-0.79}$  &  2.73$^{+0.55}_{-0.46}$  \\
NGC 4486       & 16.1  &  34.3$^{+9.7}_{-9.7}$    &  6.86$^{+1.37}_{-1.14}$  \\
NGC 4649       & 16.8  &  20.0$^{+4.0}_{-6.0}$    &  6.04$^{+1.21}_{-1.00}$  \\
NGC 4697       & 11.7  &  1.70$^{+0.2}_{-0.1}$    &  4.00$^{+0.80}_{-0.67}$  \\
NGC 5845       & 25.9  &  2.40$^{+0.4}_{-1.4}$    &  3.22$^{+0.64}_{-0.54}$  \\
NGC 6251       & 101$h^{-1}_{73}$  &  5.80$^{+1.8}_{-2.0}$  &  11.8$^{+2.36}_{-1.97}$  \\
NGC 7052       &  60$h^{-1}_{73}$  &  3.40$^{+2.4}_{-1.3}$  &  4.55$^{+0.91}_{-0.76}$  \\
%
\multicolumn{4}{c}{Bulges of Disk Galaxies} \\
Milky Way$^{b}$ & 0.008 & 0.040$^{+0.003}_{-0.003}$   & 1.32$^{+0.26}_{-0.22}$  \\
NGC  221$^{c}$  & 0.81  & 0.025$^{+0.005}_{-0.005}$   & 1.51$^{+0.30}_{-0.25}$  \\
NGC 1023        & 11.4  & 0.44$^{+0.05}_{-0.05}$      & 2.01$^{+0.40}_{-0.34}$  \\
NGC 2778$^{d}$  & 22.9  & 0.14$^{+0.08}_{-0.09}$      & 1.60$^{+0.32}_{-0.27}$  \\
NGC 2787$^{e}$  &  7.5  & 0.41$^{+0.04}_{-0.05}$      & 1.97$^{+0.39}_{-0.33}$  \\
NGC 3031        &  3.9  & 0.68$^{+0.07}_{-0.13}$      & 3.26$^{+0.65}_{-0.54}$  \\
NGC 3115$^{f}$  &  9.7  & 9.20$^{+3.00}_{-3.00}$      & 13.0$^{+2.60}_{-2.17}$  \\
NGC 3245        & 20.9  & 2.10$^{+0.05}_{-0.05}$      & 4.31$^{+0.86}_{-0.72}$  \\
NGC 3384$^{e}$  & 11.6  & 0.16$^{+0.01}_{-0.02}$      & 1.72$^{+0.34}_{-0.29}$  \\
NGC 4258$^{g}$  &  7.2  & 0.39$^{+0.01}_{-0.01}$      & 2.04$^{+0.41}_{-0.34}$  \\
NGC 4342$^{e}$  & 17.0  & 3.30$^{+1.9}_{-1.1}$        & 5.11$^{+1.02}_{-0.85}$  \\
NGC 4564$^{h}$  & 15.0  & 0.56$^{+0.03}_{-0.08}$      & 3.15$^{+0.63}_{-0.53}$  \\
NGC 7457        & 13.2  & 0.035$^{+0.011}_{-0.014}$   & 1.83$^{+0.37}_{-0.31}$  \\
\enddata
\vspace{-2mm}
\tablenotetext{a}{$B$-band image.}
\tablenotetext{b}{2.4-$\mu$m minor-axis profile from Kent, Dame, \& Fazio (1991).}
\tablenotetext{c}{Taken from Graham (2002).}
\tablenotetext{d}{NGC~2778 is a misclassified S0 galaxy (Rix, Carollo, \& Freeman 1999).}
\tablenotetext{e}{HST F814W image.}
\tablenotetext{f}{$I$-band image.}
\tablenotetext{g}{Thuan-Gunn $r$ image.}
\tablenotetext{h}{NGC~4564 is a misclassified S0 galaxy (Trujillo et al.\ 2004).}
\vspace{-2mm}
\tablecomments{
Distances are taken from Tonry et al.\ (2001, their table 1), except for the Milky Way 
(Eisenhauer et al.\ 2005), 
NGC~4342 (Virgo cluster distance modulus = 31.15, Jerjen, Binggeli, \& Barazza 2004), 
NGC~6251 ($v_{\rm CMB}$=7382 km s$^{-1}$, Wegner et al.\ 2003), and 
NGC~7052 ($v_{\rm CMB}$=4411 km s$^{-1}$, Wegner et al.\ 2003). 
These four galaxies are not listed in Tonry et al.\ (2001). 
A Hubble constant of $H_0 = 73$ km s$^{-1}$ Mpc$^{-1}$ (Blakeslee et al.\ 2002;
Spergel et al.\ 2006) has been used for the latter two galaxies. 
The SMBH masses are from the compilation in Tremaine et al.\ (2002), except for 
the Milky Way (Ghez et al.\ 2003, see also Beloborodov et al.\ 2006), 
NGC~821 (Richstone et al.\ 2006), 
NGC~3379 (Gebhardt et al.\ 2000; see also Shapiro et al.\ 2006), 
NGC~4486 (Macchetto et al.\ 1997) and 
NGC~3115 (Emsellem, Dejonghe, \& Bacon 1999).
Our sample includes a further three galaxies not listed in Tremaine et al.\ (2002).
The SMBH mass for NGC~3031 and NGC~1399 are from Merritt \& Ferrarese (2001a)
and Houghton et al.\ (2006), respectively, and the 
mass for NGC~4374 is from Maciejewski \& Binney (2001, with 
updated errors taken from Kormendy \& Gebhardt 2001). 
The S\'ersic indices are from the major-axis, light-profiles 
of $R$-band images, except where noted otherwise.
}
\end{deluxetable}


\section{The M$_{\rm bh}$--$n$ correlation}

\subsection{A linear relation}

The $M_{\rm bh}$--$n$ relation presented in Graham et al.\ (2003a) was
constructed using a {\it bisector} linear regression analysis which
treated both variables equally. 
In that study, the relation was derived with the goal of determining
the intrinsic physical relation between $M_{\rm bh}$ and the S\'ersic
shape parameter $n$ of the bulge.  Here our objective is different
because we wish to obtain a relation that can be used to {\it predict}
values of $M_{\rm bh}$ in other galaxies from their observed bulge
S\'ersic index $n$.  We therefore desire an $M_{\rm bh}$--$n$ relation
which minimizes the scatter in the quantity to be predicted, and have
thus performed an ordinary least squares (OLS) regression of $M_{\rm
bh}$ on $n$ for those local galaxies for which both quantities are
known (Table~\ref{Table1})\footnote{See Feigelson \& Babu (1992) for a
clear exposition behind the rationale of when to use what type of
regression.}.

Following Graham et al.\ (2003a), we have used a 20 per cent
measurement error
on the values of $n$, or more specifically, we have assigned an error
of $\pm\log(1.2)$ to $\log(n)$.  
Later on we explore the influence of varying this quantity. 
Several factors can contribute to the 
size of this term, including errors in the sky-subtraction,
uncertainties in the point-spread-function, the presence of bars which
are typically not modeled in bulge/disk decompositions, the influence
of additional nuclear components such as star clusters or nuclear
disks, and partially depleted cores.  The latter two issues can be
dealt with by either simultaneously fitting a S\'ersic function plus
some additional function to account for the excess flux above that of
the host galaxy (e.g., Graham \& Guzm\'an 2003; Ferrarese et al.\
2006a), or with the use of the core-S\'ersic model (Graham et al.\
2003b; Trujillo et al.\ 2004).
However, due to these various issues, it can be difficult to acquire
reliable individual uncertainties on the S\'ersic index for every
galaxy, hence our use of an average relative error.  Similarly, the
use of a fixed relative error of 13\% and 5\% was assigned to the
velocity dispersion term used by Merritt \& Ferrarese (2001b) and
Tremaine et al.\ (2002), respectively, in their construction of the
$M_{\rm bh}$--$\sigma$ relation (with the exception that they both
used a 20\% uncertainty for the Milky Way's velocity
dispersion).
%

\begin{figure*}
\includegraphics[angle=270,scale=0.62]{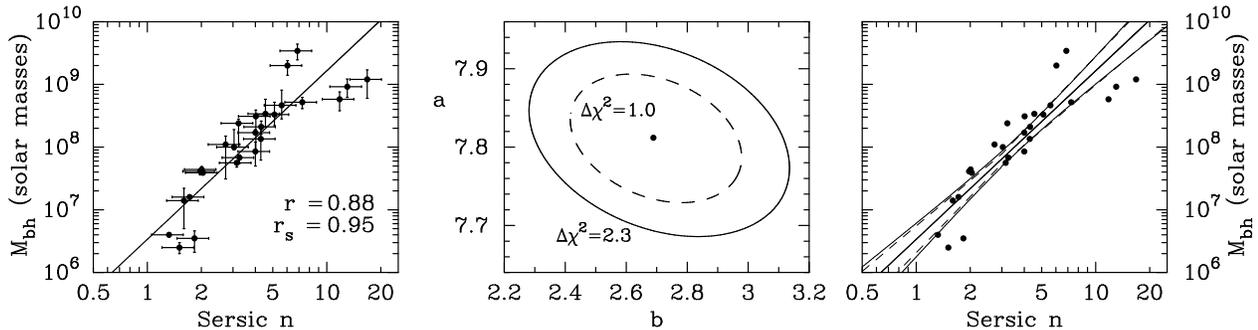}
\caption{
Correlation between a galaxy's supermassive black hole mass and the
shape parameter (i.e.\ S\'ersic index $n$) of its dynamically hot
component.  The Pearson linear correlation coefficient $r$ is given,
as is the Spearman rank-order correlation coefficient $r_s$. (The 
uncertainties on the data points were not used when computing these 
correlation coefficients.)
The regression line shown in the left panel was obtained using a
modified version (Tremaine et al.\ 2002) of the routine FITEXY (Press
et al.\ 1992, their Section~15.3), see equation~\ref{Eq_M_n}.
Consistent results were obtained using the ordinary least squares
($\log(M_{\rm bh}) \mid \log(n/3)$) linear-regression routine from
Akritas \& Bershady (1996), see equation~\ref{Eq_BCES}.
The middle panel shows the $\Delta \chi^2=1.0$ and 2.3 boundaries
around the optimal intercept, $a=7.81$, and slope, $b=2.69$.  The
projection of the $\Delta \chi^2=1.0$ ellipse onto the vertical and
horizontal axis gives the 1-$\sigma$ uncertainties $\delta a$ and
$\delta b$, respectively.  The $\Delta \chi^2=2.3$ ellipse denotes the
1-$\sigma$ {\it two-dimensional} confidence region.
This has been mapped into the right panel, and is traced by the two
solid curves.  The dashed lines in this panel are the (more commonly
used) approximations obtained using ($a\pm\delta a$) and ($b\pm\delta
b$).  The two confidence regions agree well, although the region traced 
by the dashed lines is, as expected, smaller. 
}
\label{fig-n-BH}
\end{figure*}

We have used Tremaine et al.'s (2002) modified version of the routine
FITEXY (Press et al.\ 1992), to solve the equation $y=a+bx$, by
minimising the quantity
\begin{equation}
\chi^2 = \sum_{i=1}^N \frac{( y_i- a - bx_i)^2}
    { {\delta y_i}^2 + b^2{\delta x_i}^2 + \epsilon^2 }. 
\label{SmaChi}
\end{equation}
The measurement errors on $x_i$ and $y_i$ are denoted by 
$\delta x_i$ and $\delta y_i$, and the intrinsic scatter 
$\epsilon$ is searched for by repeating the fit until
$\chi^2/(N-2)$ equals 1.  The uncertainty on $\epsilon$ is 
obtained when the reduced chi-squared value, $\chi^2/(N-2)$, 
equals $1 \pm \sqrt{2/N}$.  Doing so, one obtains 
\begin{equation}
\log (M_{\rm bh}) = 2.69(\pm 0.28)\log(n/3) + 7.81(\pm 0.08), 
\label{Eq_M_n}
\end{equation}
with $\epsilon=0.31^{+0.08}_{-0.07}$ dex in $\log M_{\rm bh}$. 
This fit is shown in Figure~\ref{fig-n-BH}. 
The {\it total} absolute scatter in $\log M_{\rm bh}$ is 0.39 dex. 
Using either an optimistic $\sim$10\% estimate for the uncertainty on 
$n$ (specifically, using $\log[n] \pm \log[1.1]$), 
or a more liberal $\sim$25\% uncertainty 
(we used $\log[n] \pm \log[1.25]$), see for example 
Caon, Capaccioli, \& D'Onofrio (1993), 
had no significant (1-$\sigma$) affect on either the
slope or intercept of the above relation.  The new intrinsic scatter
was 0.35 dex and 0.27 dex, respectively. 
%

The maximum 1-$\sigma$ error on the predicted value of $M_{\rm bh}$ in
galaxies for which $n$ is known is acquired by assuming uncorrelated
errors on $n$ and the slope and intercept of the $M_{\rm bh}$--$n$
relation.
Gaussian error propagation for the linear equation 
$y=(b\pm\delta b)(x\pm \delta x) + (a\pm \delta a)$, gives an 
error on $y$ equal to
\begin{eqnarray}
\delta y & = & \nonumber \\ 
 & & \hskip-40pt \sqrt{ (dy/db)^2(\delta b)^2 + (dy/da)^2(\delta a)^2 + (dy/dx)^2(\delta x)^2 } \nonumber \\
 & = & \sqrt{ x^2(\delta b)^2 + (\delta a)^2 + b^2(\delta x)^2 }. \nonumber
\end{eqnarray}
In the presence of intrinsic variance, $\epsilon$, 
the uncertainty on $y$ will be greater, such that, assuming the
intrinsic variance is in the $y$ coordinate, 
\begin{eqnarray}
\delta y = \sqrt{ x^2(\delta b)^2 + (\delta a)^2 + b^2(\delta x)^2 + \epsilon^2}. \nonumber
\end{eqnarray}
For our expression (equation~\ref{Eq_M_n}) we have
$x=\log(n/3)$, and so $dx/dn = 1/[\ln(10).n]$, and therefore 
\begin{eqnarray}
(\delta \log M_{\rm bh})^2 & = &
 [\log(n/3)]^2(0.28)^2 + (0.08)^2 \nonumber \\
 & & \hskip-40pt + [2.69/\ln(10)]^2 [\delta n/n]^2 + (0.31)^2. 
\label{EqMerr}
\end{eqnarray} 

Given the debate over the slope of the $M_{\rm bh}$--$\sigma$
relation (Merritt \& Ferrarese 2001b; Tremaine et al.\ 2002; Novak et al.\ 2006), 
it is of interest to know how much the uncertainty in the
slope and intercept of the $M_{\rm bh}$--$n$ relation may contribute
to the uncertainty in the predicted SMBH masses.  From
equation~\ref{EqMerr}, if one measures a bulge to have $n=3$ with 
$\delta n/n=0.2$, then the error in $n$ contributes 96\% of
$\delta \log M_{\rm bh}$.  That is, the uncertainty on the slope and
intercept of equation~\ref{Eq_M_n} are not substantial contributors to
the error budget on $\log M_{\rm bh}$.  
If one has a galaxy with $n=1$ or 9, and again $\delta n/n=0.2$, 
then the combined error from the uncertainty on the slope and
intercept contributes only 14\% of the error on $\log M_{\rm bh}$.

We have additionally used the OLS regression analysis BCES($\log
M_{\rm bh} \mid \log n$) from the code of Akritas \& Bershady (1996),
which allows for both measurement errors and intrinsic scatter. 
For our sample of $N=27$ galaxies, we obtained
\begin{equation}
\log (M_{\rm bh}) = 2.68(\pm 0.40)\log(n/3) + 7.82(\pm 0.07), 
\label{Eq_BCES}
\end{equation}
with $\epsilon=0.30^{+0.09}_{-0.07}$ dex in $\log M_{\rm bh}$. 
Using an uncertainty of $\pm \log(1.25)$ or $\pm \log(1.1)$
for the value of $\log(n)$ had no significant (1$\sigma$) affect
on either the slope or intercept of the above relation --- which agrees 
well with that in equation~\ref{Eq_M_n}.
%

\subsubsection{Symmetrical regression} 

To obtain the intrinsic astrophysical relation, the modified 
version of FITEXY which minimizes equation~\ref{SmaChi} 
should not be used. The reason is because it is 
biased --- to produce a
low slope --- by the minimisation of the intrinsic variance, $\epsilon$, 
along the $y$ (i.e.\ $\log M_{\rm bh}$) axis.  If the minimisation
is instead performed along the $x$ (i.e.\ $\log n$) axis, then the
$\epsilon^2$ term in the denominator of equation~\ref{SmaChi} will be 
replaced with $b^2\epsilon^2$.  
Making this substitution, and performing the new regression\footnote{We 
have ignored unknown, but possible,
selection boundaries in $\log n$ that could bias such a fit (see, e.g.,
Lynden-Bell et al.\ 1988, their Figure~10).}, 
one obtains $b=3.10\pm0.33, a=7.78\pm0.09$, and 
$\epsilon=0.11^{+0.04}_{-0.02}$ dex in $\log n$.

The average of the above two slopes from the modified FITEXY routine, 
(3.10+2.69)/2 = 2.90, agrees well with the slope obtained using the 
{\it bisector} linear regression routine BCES from Akritas \& Bershady
(1996), which gives
\begin{equation} 
\log (M_{\rm bh}) = 2.85(\pm 0.40)\log(n/3) + 7.80(\pm 0.07). 
\label{Eq_Sym}
\end{equation}
This relation is in good agreement with the
(intrinsic astrophysical relation) presented in Graham et al.\ (2003a). 
%

\subsection{A curved relation}

As noted in Graham et al.\ (2001, 2003a), we have no {\it a priori}
knowledge that the $M_{\rm bh}$--$n$ relation is linear.  For this
reason, in those papers we employed the use of both linear and
non-linear statistics to measure the correlation strength.  Here we go
one step further by fitting a quadratic to the data.

We stress that, from a physical stand point, we do not know what the
actual form of the relation should be.  The quadratic equation which
we adopt is an empirical model.  In a Taylor series expansion it is
simply the next order term.  The $M_{\rm bh}$--$n$ data may in fact be
described by a double power-law, however this would require the use of
{\it four} free parameters (a low- and high-mass slope, and a
transition mass and transition S\'ersic index).  The quadratic
relation has only three parameters and is the adopted model for 
explorations of non-linearity in the $M_{\rm bh}$--$\sigma$ data 
(Wyithe 2006a,b).

In passing we note that there have been claims that the $M_{\rm 
bh}$--$\sigma$ relation may not be linear, but has either negative
curvature (e.g., Granato et al.\ 2004; Cirasuolo et al.\ 2005)
positive curvature (e.g. Hopkins \& Hernquist 2006), no
curvature at even the 0.75-1.5 $\sigma$ level (Wyithe 
2006b)\footnote{The tighter constraint of 0.75 $\sigma$ (that is, 
the factor in front of a quadratic term is inconsistent with zero 
at only the 0.75 $\sigma$ level), 
comes from using SMBHs with resolved sphere's of influence.
Including galaxies with unresolved sphere's of influence
and adding in single epoch reverberation mapping masses (uncertain
to factors of 3-4) weakens the result of a 
purely linear $M_{\rm bh}$--$\sigma$ relation, with the
probability that the second order term does not equal zero
ruled out at the (still weak) 1-5-2 $\sigma$ level.}, 
or curves down at low SMBH masses and up at high
SMBH masses (Sazonov et al.\ 2005).  Given that the $L$--$\sigma$
relation is not linear, having a slope of $\sim$4 at the bright end
and $\sim$2 at the faint end (Tonry 1981; Davies et al.\ 1983; Held et
al.\ 1992; De Rijcke et al.\ 2005; Matkovi\'c \& Guzm\'an 2005), and
{\it if} the $M_{\rm bh}$--$L$ relation {\it is} linear, then the
$M_{\rm bh}$--$\sigma$ relation obviously cannot be linear, but must
have a positive curvature.  Of course, the $M_{\rm bh}$--$L$ relation
may not be linear.
%

\begin{figure*}
\includegraphics[angle=270,scale=0.64]{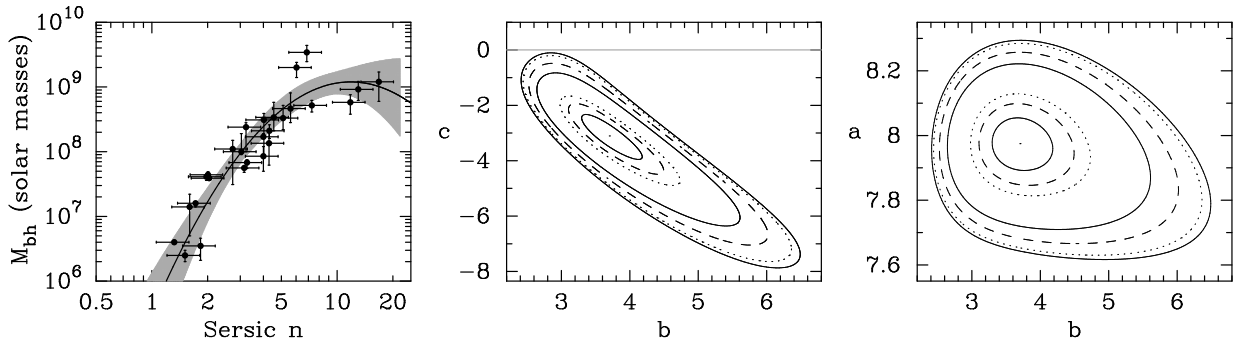}
\caption{
Left panel: The data points are the same as in Figure~\ref{fig-n-BH}, 
but this time the optimal 3-parameter 
log-quadratic relation (equation~\ref{EqQuad}) is shown.  The absolute
scatter in $\log M_{\rm bh}$ is 0.31 dex. 
The shaded area corresponds to the $\Delta \chi^2 =3.53$ 
(1-$\sigma$) confidence region. 
Middle and right panel:  
The inner, central and outer solid contours denote the $\Delta \chi^2
=1.00$, 9.00 and 15.1 boundaries, respectively: their projections onto
the vertical and horizontal axes give the 68.3\% (1-$\sigma$)
99.73\% (3-$\sigma$) and 99.99\% confidence intervals on
the individual parameters in equation~\ref{EqQuad}.
The inner and outer dashed lines denote the $\Delta \chi^2 =2.30$ and
11.8 boundaries, respectively, delineating the 1-$\sigma$ and
3-$\sigma$ two-dimensional confidence regions.
The dotted lines show the $\Delta \chi^2 =3.53$ and 14.2 boundaries,
respectively.  They correspond to the 1-$\sigma$ and 3-$\sigma$
three-dimensional confidence ellipsoids.  (see Press et al.\ 1992,
their Section~15.6).  
}
\label{Fig-quad}
\end{figure*}

We fit the quadratic equation $y = a + bx + cx^2$, to the ($M_{\rm
bh},n$) data by minimising the statistic
\begin{equation}
\chi^2 = \sum_{i=1}^N \frac{( y_i - a - bx_i - cx_i^2)^2}
    { {\delta y_i}^2 + (2cx_i + b)^2{\delta x_i}^2
       + \epsilon^2 }, 
\label{BigChi} 
\end{equation}
where $y_i= \log(M_{{\rm bh},i})$, $x_i=\log(n_i/3)$, and $\epsilon$
is the intrinsic variance which, given our objective of predicting new
SMBH masses, we attribute entirely to reside in 
the $y$ $(\log M_{\rm bh})$ direction.  
Solving for $\chi^2/(N-3)=1$, we find 
\begin{eqnarray}
\log (M_{\rm bh}) & = & 7.98(\pm0.09) +  3.70(\pm 0.46)\log(n/3) \nonumber \\
& & -3.10(\pm 0.84)[\log(n/3)]^2, 
\label{EqQuad}
\end{eqnarray}
with $\epsilon = 0.18^{+0.07}_{-0.06}$ dex. 
%
The total absolute scatter in $\log M_{\rm bh}$ is 0.31 dex. 
From Figure~\ref{Fig-quad}, the term $c$ is inconsistent
with a value of zero at greater than the 99.99\% confidence 
level\footnote{Excluding 
the two galaxies with the highest SMBH masses, the total absolute scatter
reduces to 0.25 dex, the intrinsic scatter drops to zero, and one finds that 
$a=7.90, b=3.22$, and $c=-2.55$, only a 1-$\sigma$ deviation from the 
values obtained using the full data set.  Using the full data set, but with the 
intrinsic scatter set to zero, the value of $c$ is still inconsistent with 
a value of zero at the 3-$\sigma$ level.}.

With small data sets, formal errors can under-estimate the true error
on a fitted parameter.  To explore the danger that we may be
under-estimating the uncertainty on the parameter $c$ in front of the
quadratic term in equation~\ref{EqQuad}, we use Bootstrap Monte
Carlo simulations.  The random number generator {\bf ran1} from Press
et al.\ (2002) was used in a bootstrapping process that involved
sampling (with replacement) from the original 27 data points.  One
thousand data sets containing 27 data points each were then
individually fitted to find the optimal log-quadratic equation,
exactly as done with the original data.  This then gave 1000 new
estimates for each of the three parameters ($a, b, c$).  The range in
values covered by the central 68.3\% of these three data sets gives one 
an estimate of the 1-$\sigma$ confidence intervals --- without 
making any assumption about (the Gaussianity of) the distribution.  
This parameter range is found to be (7.88...8.04, 3.26...4.24, $-$3.98...$-$2.30). 
This is in good agreement with the 1-$\sigma$ 
parameter uncertainties obtained from Figure~\ref{Fig-quad} and given
in equation~\ref{EqQuad}.  We therefore conclude that we are not
under-estimating our errors, and that the $M_{\rm bh}$--$n$ relation
is indeed curved.

Using equation~\ref{EqQuad}, and the value of $n$ in other galaxies, 
the maximum error on the predicted value $\log M_{\rm bh}$, denoted by
$\delta \log M_{\rm bh}$, can be written as
\begin{eqnarray}
(\delta y)^2  =  x^4(\delta c)^2 + x^2(\delta b)^2 + 
(\delta a)^2 \nonumber \\ 
+ (2cx+b)^2(\delta x)^2 + \epsilon^2, \nonumber
\end{eqnarray}
which equates to 
\begin{eqnarray}
(\delta \log M_{\rm bh})^2 & = & [\log(n/3)]^4 +[\log(n/3)]^2/4 + (0.09)^2 \nonumber \\
 & & \hskip-75pt +[3.70-6.20\log(n/3)]^2 [\delta n/n]^2/[\ln(10)]^2 + (0.18)^2. 
\label{errQuad}
\end{eqnarray}
Assuming $\delta n/n = 0.2$, Table~\ref{Table2} gives the value of
$\delta \log M_{\rm bh}$ for different values of $n$.  Obviously the
error in the slope of the relation (similarly for the linear relation)
causes the masses to be less well constrained at the end of the
relation.  Due to the steepness of the quadratic relation at low $n$,
the uncertainty on the estimated SMBH masses can be large there.
%


In passing we note one of the implications of a curved
$M_{\rm bh}$--$n$ relation.  Either the $M_{\rm bulge}$--$n$ 
relation must be curved, 
or the $M_{\rm bh}/M_{\rm bulge}$ ratio can not be constant with mass. 

\begin{deluxetable}{lccccccc}
\tablewidth{0pt}
\tablecaption{Black hole mass uncertainties\label{Table2}}
\tablehead{
\colhead{$n=$} & \colhead{0.5}  & \colhead{1.0}  & \colhead{2.0}  & 
\colhead{3.0}  &  \colhead{4.0}  &  \colhead{8.0}  &  \colhead{10.0} 
}
\startdata
\multicolumn{8}{c}{Linear Fit (equation~\ref{Eq_M_n})} \\
       $\log M_{\rm bh}$                 & 5.72 & 6.53 & 7.34 & 7.81 & 8.15 & 8.96 & 9.22  \\
$\delta \log M_{\rm bh}$                 & 0.45 & 0.42 & 0.40 & 0.40 & 0.40 & 0.41 & 0.42  \\
                                       &      &      &      &      &      &      &         \\
\multicolumn{8}{c}{Quadratic Fit (equation~\ref{EqQuad})} \\
       $\log M_{\rm bh}$                 & 3.22 & 5.51 & 7.23 & 7.98 & 8.39 & 8.99 & 9.07  \\
$\delta \log M_{\rm bh}$                 & 1.05 & 0.70 & 0.47 & 0.38 & 0.33 & 0.36 & 0.43  \\
\enddata
\tablecomments{
Uncertainty, $\delta \log M_{\rm bh}$ dex, on an estimated black hole
mass using the linear and quadratic $M_{\rm bh}$--$n$ relations as a
function of $n$.  The steepness of the quadratic relation at low $n$
results in both a better fit to the data for $ 1 \le n \le 2$, but
also an increased uncertainty on the predicted value of $\log M_{\rm
bh}$.
}
\end{deluxetable}

\subsection{The high-mass end}

An obvious consequence of our second order, log-quadratic, relation is
the different expected masses at the ends of the relation.  For
example, at the high mass end, we no longer predict infinitely large
SMBHs.  Obviously a linear $M_{\rm bh}$--$n$ relation or a linear
$M_{\rm bh}$--$\sigma$ relation, and a positively curving
log-quadratic $M_{\rm bh}$--$\sigma$ relation, imply infinitely
massive SMBH masses at the high-$\sigma$ end.  In contrast, our
negatively curving $M_{\rm bh}$--$n$ relation suggests a maximum mass
limit to which the universe has built SMBHs\footnote{We do not equate this 
with a upper mass to which the universe {\it can} build SMBHS,
only with what the universe {\it has} built.}. The maximum mass occurs
where the derivative of equation~\ref{EqQuad} equals zero, i.e., where
$d\log(M_{\rm bh})/d\log(n/3) = 2cx+b = 2(-3.10)\log(n/3)+3.70 = 0$.
The value of $n$ at this rather broad peak and turnover is 11.9.
Scanning the $\Delta \chi^2 = 3.53$ (1-$\sigma$) confidence region shown in
Figure~\ref{Fig-quad}, so as to provide error bounds, this peak
corresponds to a maximum SMBH mass of $1.2^{+2.6}_{-0.4}\times10^9
M_{\sun}$.  This upper value matches well with the uppermost mass
reported in several observational studies (e.g., McLure et al.\ 2004;
Xie, Zhou, \& Liang 2004; Pian, Falomo, \& Treves 2005; Sulentic et al.\ 2006;
Kollmeier et al.\ 2006)
and intriguingly with the upper mass limit for axion bubbles (Svidzinsky 2006). 
We note that this peak is reached somewhat asymptotically as the value of
$n$ increases, and that we do not expect galaxies with larger values of
$n$ to have significantly smaller SMBH masses than the peak mass.  

We additionally note that some studies of high-redshift quasars have
reported SMBH masses up to $10^{10} M_{\sun}$.  If correct, such
objects would represent a challenge for the quadratic relation defined
by the current local sample.  Of course, to properly solve this
discrepancy requires {\it direct} SMBH mass measurements.  With the
upcoming ten-fold increase in spatial resolution (compared to the
Hubble Space Telescope), the next generation of extremely large
telescopes, such as the Giant Magellen Telescope, will be able to
resolve the sphere of influence around SMBHs in galaxies ten times
further away.  The resultant 1000-fold increase in survey volume
should help address the issue of just how massive SMBHs are. 

It is also worth remarking that the upper mass limit we find is
somewhat of a consequence of our use of a quadratic relation.  That
is, had we used a four-parameter double power-law, we would not
observe such behaviour.  Although, Figure~\ref{Fig-quad} does suggest
that some kind of asymptotic function, which reaches a maximum SMBH
mass, would also provide a good description to the data.  We will,
however, postpone a comparison of assorted arbitrary functions for a
later time.

The two galaxies with the highest SMBH masses (NGC~4486 and 4649)
appear as outliers from the curved relation. 
We have checked their S\'ersic indices are correct and it may be of value to
re-examine their black hole masses.  While this is beyond the scope of
the present paper, we note that there is reason to suspect that
NGC~4486 (M87) might have had its black hole mass over-estimated.
Maciejewski \& Binney (2001) have shown that the SMBH mass in NGC~4374
was reduced by a factor of nearly four, from $1.7\times10^9 M_{\sun}$
to $4.64\times10^8 M_{\sun}$, after the slit-width was appropriately
dealt with.  NGC~4486, like NGC~4374, has similarly had its SMBH mass
derived from the emissions of a circumnuclear gaseous disk using
slit-spectroscopy.  
These galaxies also appear to have overly large SMBH masses when compared
with the theoretical models of Menci et al.\ (2006, their Figure~1). 
It may be advantageous to confirm this galaxy's
SMBH mass using AO-assisted integral field spectroscopy with an
instruments such as NIFS (McGregor et al.\ 2003) or SINFONI (Eisenhauer
et al.\ 2003).  
%


\subsection{The low-mass end}

%
Our negatively curving $M_{\rm bh}$--$n$ relation suggests that the SMBHs 
in dwarf elliptical galaxies will be smaller than predicted from the
linear $M_{\rm bh}$--$n$ relation.  
When $n=1$ ($M_B\sim -14.5\pm1.5$ mag; 
Graham \& Guzm\'an 2003, their figure~10), one can expect a SMBH mass of
$0.32^{+0.68}_{-0.24}\times10^6 M_{\sun}$ and 
$3.4^{+4.1}_{-1.9}\times10^6 M_{\sun}$ using 
the quadratic and linear relations, respectively.  
At $n=0.5$
%
%
one would expect a SMBH mass of $1.6^{+16.8}_{-1.5}\times10^3 M_{\sun}$ and
$5.2^{+7.5}_{-3.1}\times10^5 M_{\sun}$ from the quadratic and linear relations,
respectively.  
When $n=0.25$, the (extrapolated) quadratic relation
predicts a mass of only $\sim 2M_{\sun}$.  The absence of 10$^5 M_{\sun}$ 
black hole detections in local dwarf galaxies (e.g., Valluri et al.\
2005) and M33 (Gebhardt et al.\ 2001; Merritt, Ferrarese, \& Joseph 2001) 
would argue against the extrapolation of the linear relation. 
The suitability of the curved log-quadratic $M_{\rm bh}$--$n$ relation
to globular clusters (e.g., Gebhardt, Rich, \& Ho 2005; De Rijcke,
Buyle, \& Dejonghe 2006, and references therein) remains to be
explored.

\section{Central star clusters}

\begin{figure}
\includegraphics[angle=270,scale=0.56]{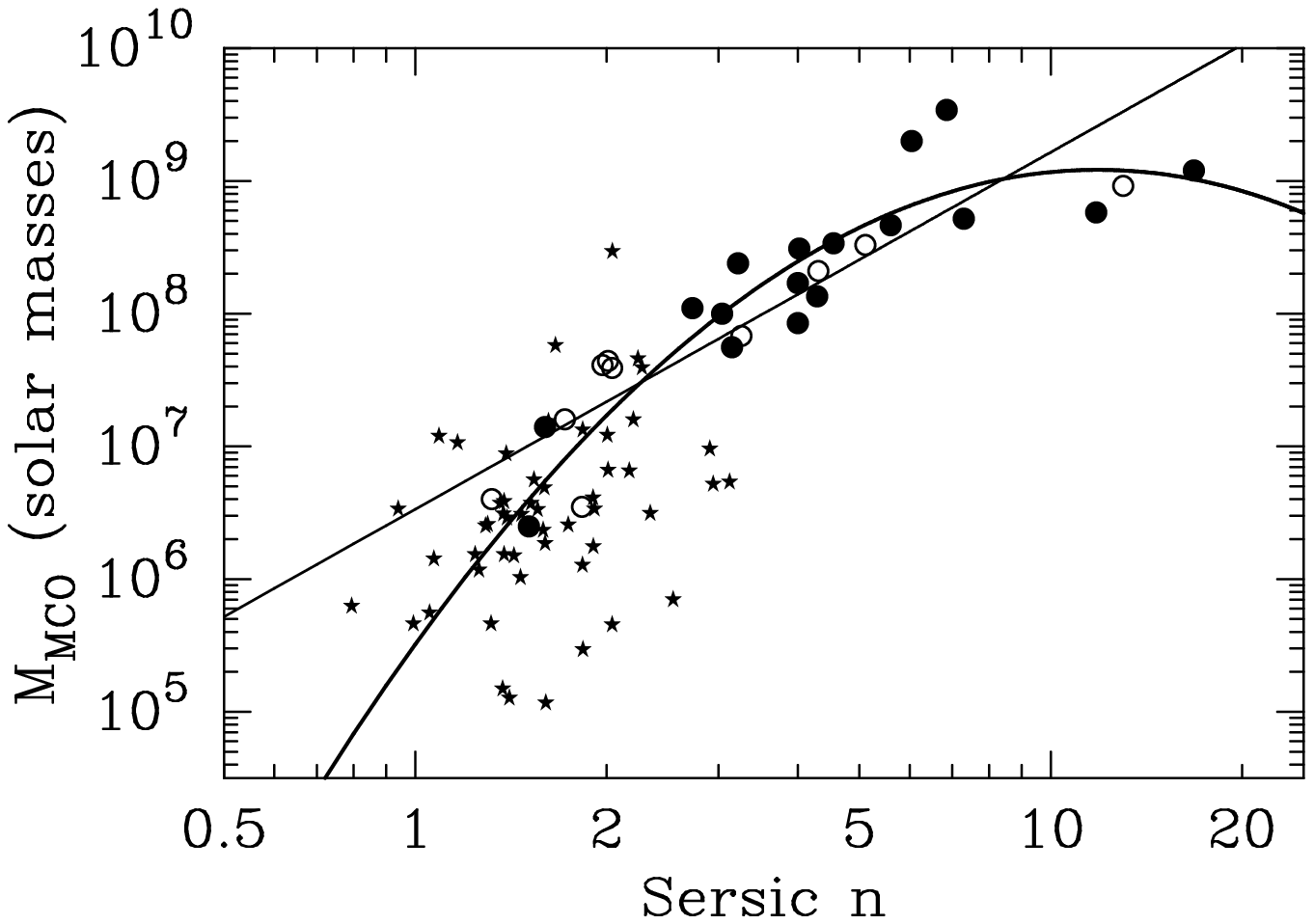}
\caption{
Central `massive compact object' mass versus host
galaxy S\'ersic index `$n$'.  Filled circles represent SMBHs
in elliptical galaxies, while open circles represent SMBHs
in disk galaxies. 
The stars represent nuclear star clusters in elliptical
galaxies presented in 
Graham \& Guzm\'an (2003) and Ferrarese et al.\ (2006).  
The straight line is the same as that shown in Figure~\ref{fig-n-BH}, and the 
curved line is the same as that shown in Figure~\ref{Fig-quad}. 
}
\label{Fig-MCO}
\end{figure}

From an analysis of HST images of low-luminosity elliptical galaxies
in the Coma cluster, Graham \& Guzm\'an (2003) presented a correlation
between the luminosities of nuclear star clusters and that of their host
galaxy.  Similarly, strong correlations involving Virgo
cluster ellipticals have been reported (e.g. C\^ot\'e et al.\ 2006) and
also between the nuclear star clusters in the bulges of spiral
galaxies and their host bulge's flux (e.g., Balcells et al.\ 2003).
The cluster-to-host flux ratio reported in Graham \&
Guzman (2003) was a few tenths of one per cent, intriguingly comparable
to the mass ratio observed between SMBHs and their host bulge.
In Figure~\ref{Fig-MCO} we have expanded the $M_{\rm bh}$--$n$ diagram
by including the masses of nuclear star clusters plotted against their
host galaxy's S\'ersic index.  We have used two data sets obtained
with the Hubble Space Telescope, and therefore likely to have had a
reliable bulge/star-cluster decomposition.

The first data set is from Graham \& Guzm\'an (2003, their Table~2).
We have converted their F606W nuclear star cluster magnitudes into masses
using an F606W-V color of $-$0.3 mag (Fukugita, Shimasaku \&
Ichikawa 1995) and a $V$-band mass-to-light ratio $M/L=1.5 (M_{\sun}/L_{\sun})$.  
This ratio is the median of the (King model) dynamical $M/L$ ratios from 57
star clusters listed in McLaughlin \& van der Marel (2005, their
Table~13; see also Pryor \& Meylan 1993, their Table~2).  The second
data set comes from Ferrarese et al.\ (2006a, their Table~3), for
which we have used an F475W-V color of 0.2 mag and $M/L_V=1.5 (M_{\sun}/L_{\sun})$. 
Due to the association of SMBHs with their host bulge, rather than the
host galaxy, we have exluded the lenticular
galaxies\footnote{Inclusion of the lenticular galaxies increases the
scatter but does not suggest any shift in one particular direction.}
because Ferrarese et al.\ (2006a) modeled the combined bulge$+$disc
light with a single S\'ersic function.  That is, no bulge/disc
decomposition was performed.


Figure~\ref{Fig-MCO} reveals (albeit with some scatter which is in
part likely due to our conversion from flux to mass) that the nuclear
star clusters appear to follow the same log-quadratic relation as
defined by the ($M_{\rm bh},n$) data set.
Clearly there is no sharp transition; both SMBHs and star clusters
have overlapping masses in the range $3 \times 10^6 < M_{\rm
mco}/M_{\sun} < 10^8$.  Furthermore, the lower limit to the SMBH
masses may be due to selection effects, i.e., a reflection of our
inability to resolve the SMBH sphere of influence in low-mass bulges
(Merritt \& Ferrarese 2001a).  Similar results using host bulge mass,
rather than S\'ersic index, have recently been reported in C\^ot\'e et
al.\ (2006), Ferrarese et al.\ (2006b), Rossa et al.\ (2006) and
Wehner \& Harris (2006).  McLaughlin, King, \& Nayakshin (2006)
propose that feedback from supernovae and stellar winds from a nuclear
star-cluster can regulate the growth of the host bulge and thereby
explain the latter connection.

The result shown in Figure~\ref{Fig-MCO} is intriguing because it
suggests that the formation mechanisms of nuclear star clusters may
share some common processes with the formation of SMBHs.  The
shallower potential wells, lower central stellar densities (prior to
core-depletion in massive elliptical galxies) and shallower central
gradients in lower $n$ bulges (Terzi\'c \& Graham 2005) may result in
the production of star clusters rather than SMBHs.

An obvious question is: Do any of the galaxies in Table~\ref{Table1}
possess nuclear star clusters in addition to their SMBHs?  
For the Milky Way the answer is yes (Merritt 2006 and references therein). 
For the local group galaxy NGC~221 (M32), 
the excess central flux above the S\'ersic model may not 
represent a distinct additional central massive object, 
as is the case with the nuclei in NGC~205 (Valluri et al.\ 2005). 
In many galaxies, the presence of non-thermal flux from AGN can 
complicate the identification of
nuclear star clusters when using only images and light-profiles.  One
additionally requires spectral information.  In general, 
core-S\'ersic galaxies do not show any excess nuclear flux, and the
few which do show excess flux have an AGN.  In contrast, the majority
of the lower-luminosity S\'ersic (``power-law'') galaxies in Table~\ref{Table1} 
{\it do}
show excess nuclear flux.  However, this is usually due to an
AGN.  Two exceptions are NGC~3384 and NGC~7457, which possess both a
SMBH {\it and} a central star cluster (Ravindranath et al.\ 2001).

While the mass of the star cluster in NGC~3384 is approximately twice 
its SMBH mass, the star cluster in NGC~7457 is an order
of magnitude greater than its SMBH mass\footnote{The
apparent $H$-band magnitudes were converted into absolute magnitudes
using the distances in Table~\ref{Table1}, and then into units of
solar luminosity using $M_{\sun,H}$=3.32 (Bessell, Castelli, \& Plez
1998, their Appendix~C and D), and finally solar mass using 
$M/L_H=1.0 (M_{\sun}/L_{\sun})$.}
%
%
Creating a new $M_{\rm bh + mco}$--$n$ diagram would move NGC~7457
(which has the second lowest SMBH mass in our sample) so that it
overlaps with a cluster of three other data points seen in
Figure~\ref{Fig-quad} --- leaving it consistent with the quadratic
shown there. 

Successive mergers of nucleated, low-$n$,
low-luminosity galaxies will eventually result in a galaxy whose mass
is sufficient to expect a SMBH.  This then raises the tantalising
possibility of the {\it recent} formation of some SMBHs from the
merger of two, or more, nucleated galaxies and the subsequent merger
and `runaway collision' of their respective nuclear star clusters.  
To date, such a collisional process in star clusters has only been
invoked to explain the formation of intermediate-mass black holes
(IMBHs: Quinlan \& Shapiro 1990; Portegies Zwart et al.\ 1999;
Freitag, G\"urkan, \& Rasio 2006).
%
%
We are, however, unaware of the actual dwarf-dwarf galaxy collision
rate today, and do not wish to suggest it is a frequent phenomenon.

The formation scenario proposed here for potential {\it young} SMBHs
is perhaps, at face value, contrary to the conventional scenario in
which they form via gas accretion during the height of quasar
activity.  Indeed, massive black holes
have long been known to exist at high-redshifts, $z>3$. 
Nonetheless, we do not rule out a kind of ``downsizing'' or late-time
formation of low-mass SMBHs built through a relatively dry runaway
merger of nuclear star clusters.  While involving no significant AGN
feedback, such a process could still maintain the $M_{\rm mco}/M_{\rm
bulge}$ mass ratio.

{\it If} there are appreciable numbers of {\it young} SMBHs built from
the merger of nuclei, this may require a downward revision to
measurements of the average radiative efficiency of SMBHs and thus
their implied rate of rotation (e.g., Ferrarese 2002; Shankar et al.\
2004; Wang et al.\ 2006).

\section{Dependence on $H_0$}

The above linear and curved relations are nearly, but not completely,
independent of the Hubble constant.  While the S\'ersic indices are
independent of galaxy distance, the SMBH masses depend linearly on the
distance to each galaxy.  Distances for 24 of our 27 galaxies were
obtained from surface brightness fluctuation measurements and are thus
independent of the Hubble constant, as of course is the Milky Way center. 
However, 
a Hubble constant of 73 km s$^{-1}$ Mpc$^{-1}$ was used to derive the
distances to NGC~6251 and NGC~7052 from their redshift.

Studies which use the $M_{\rm bh}$--$n$ relation, or indeed the
$M_{\rm bh}$--$\sigma$ relation, need to know if the SMBH masses
predicted from the relation are dependent on $H_0$.  If they are, then
the Hubble constant needs to factored into such mass derivations and
thus also estimates of the SMBH mass function and space density.  We
have therefore rederived the quadratic $M_{\rm bh}$--$n$
relation excluding NGC~6251 and NGC~7052, so as to remove any 
possible dependency on $H_0$. 
%
%
Doing so, we obtain $\log (M_{\rm bh}) = 7.97(\pm0.09) + 3.75(\pm
0.46) \log(n/3) -2.92(\pm 0.83)[\log(n/3)]^2$, with $\epsilon =
0.17^{+0.09}_{-0.06}$ dex.

This is not significantly different from the
expression given in equation~\ref{EqQuad}. 
In fact, the main reason for the slight shift in 
numbers is because of the exclusion of NGC~6251 --- which resides at the 
high-$n$ end of the relation, and thus has more weight than individual 
points in the middle of the relation --- rather than the use of 
$H_0=73$ km s$^{-1}$ Mpc$^{-1}$ for this galaxy. 
This is evidenced by using $H_0=100$ km s$^{-1}$ Mpc$^{-1}$ for
NGC~6251 and NGC~7052 and adjusting their SMBH masses accordingly, 
which results in $\log (M_{\rm bh}) = 7.97 + 3.64 \log(n/3) - 3.11[\log(n/3)]^2$.
This expression is closer to equation~\ref{EqQuad} than
the expression in the previous paragraph is to equation~\ref{EqQuad}, and reveals that
the exclusion of NGC~6251 and NGC~7052 introduces more of a change
than the use of an uncertain Hubble constant.  It therefore makes
sense to use equation~\ref{EqQuad} and, given the uncertainties on the
fitted parameters $a, b$ and $c$, treat equation~\ref{EqQuad} as if 
it were independent of the Hubble constant.  The same statement
is true for equations~\ref{Eq_M_n}, \ref{Eq_BCES} and \ref{Eq_Sym}.

\section{An $M_{\rm bh}$--$\mu_0$ relation\label{Secmu0}}

There are now several relations between SMBH mass and the properties
of the host bulge, such as luminosity, mass, velocity dispersion
within some radius, and concentration.  Which of these is the most
fundamental remains unanswered (Novak et al.\ 2006).  It may be that
none of these are the drivers of the SMBH-bulge connection, but that
all are symptomatic of a more profound relation.  For instance, the
combination of concentration and central density might be what is
important --- as this defines both the baryonic distribution and the
strength of the gravitational potential with radius, at least until
dark matter becomes a significant factor.  It is therefore not
unreasonable to expect that this might influence, or possibly even
dictate, the ability of a bulge to fuel any central SMBH.

The motion of the stars within a bulge, traced through the observable
$\sigma$, is the dynamical {\it response} to the underlying
mass distribution.  While $\sigma$ is therefore a tracer of mass, it
is obviously dependent on how that mass is distributed, and is thus a) a
function of the central stellar density and b) has a radial
dependence which is set by the stellar concentration, 
at least within $\sim 1R_{\rm e}$ (Lintott, Ferreras, \& Lahav 2006).

While the SMBH mass might be a product of the radial mass distribution,
specified by the central density and concentration, the observed
relation between central density and concentration (e.g., 
Graham \& Guzm\'an 2003, their Figure~9; Merritt 2006, his Figure~5) 
subsequently means
that one can express the SMBH mass in terms of just one of these
quantities.
Extrapolating under the nuclear star clusters in low-luminosity
bulges, and over the partially-depleted cores in giant galaxies ---
whose central stellar density has been modified by the slingshot
effect of coalescing SMBHs --- gives an estimate of a bulge's
(original) central stellar density.
%
%
Given the $M_{\rm bh}$--$n$ relations derived in this paper, we can
predict a correlation between $M_{\rm bh}$ and central stellar
density.

Substituting $\log(n) = (22.8 - \mu_{0,B})/14$, where $\mu_{0,B}$ is
the central $B$-band surface density (Graham \& Guzm\'an
2003)\footnote{Many of our images are photometrically uncalibrated, so
we are unable to plot $\log(M_{\rm bh})$ versus $\mu_0$ with the
present data set.}, into equation~\ref{Eq_M_n} gives 
\begin{equation}
\log(M_{\rm bh}) = 10.91 - 0.19\mu_{0,B}.
\label{EqMuLin}
\end{equation}
A relatively large uncertainty on the central surface brightness of 1
mag arcsec$^{-2}$ translates to an uncertainty of only 0.19 dex in the
logarithm of the SMBH mass, a value comparable to the intrinsic scatter found
for equation~\ref{EqQuad}.
Substitution of the above term for $\log(n)$ into
equation~\ref{EqQuad} gives 
\begin{equation}
\log(M_{\rm bh}) = 8.13 + 0.25\mu_{0,B} - 0.016(\mu_{0,B})^2.
\label{EqMuQuad}
\end{equation}

\begin{figure}
\includegraphics[angle=270,scale=0.56]{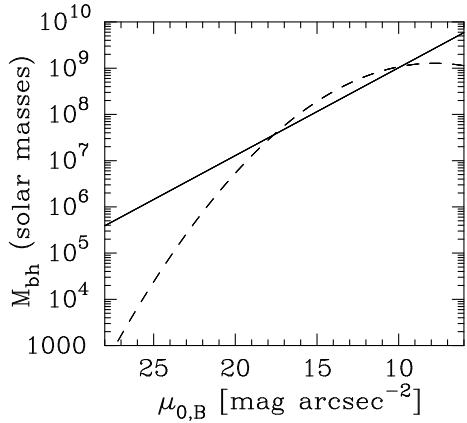}
\caption{
Predicted correlation between SMBH mass and the central $B$-band 
surface brightness of the host bulge, after allowing for 
excess nuclear flux and partially-depleted cores, as mentioned
in Section~\ref{Secmu0}.  The solid line corresponds to 
equation~\ref{EqMuLin}, the dashed line corresponds to 
equation~\ref{EqMuQuad}. 
}
\label{Fig-M-mu0}
\end{figure}

These two expressions differ most in their prediction of SMBH masses 
at the low mass end (Figure~\ref{Fig-M-mu0}), with the linear relation 
predicting $\sim10^6 M_{\sun}$ masses if $\mu_0 = 25 B$-mag, and the 
quadratic relation predicting masses of $\sim10^4 M_{\sun}$ at this
faint central surface brightness. 

Although we are unable to conclude that an $M_{\rm bh}$--$\mu_0$
relation is more, or even equally, fundamental than any other relation
involving SMBH mass, we do consider it prudent to explore this
possibility.  We hope to acquire calibrated, high-resolution,
near-infrared images to further investigate this proposed relation.
Due to the random projection angles of triaxial bulges on the plane of
the sky, one might expect a certain amount of variability in the
observed values of $\mu_{0,B}$.  We therefore note that it may be
advantageous to construct a plot of SMBH versus the deprojected
(internal) central density.  Such a program, however, is beyond the
intended scope of this paper.

\section{Summary}

We have constructed the $M_{\rm bh}$--$n$ relation using updated
SMBH masses and refined galaxy light-profile shapes, $n$.  We performed an
ordinary least squares, linear regression of $M_{\rm bh}$ on $n$ that allows
for measurement errors and intrinsic scatter $\epsilon$, to obtain   
$\log (M_{\rm bh}) =  7.81(\pm 0.08) + 2.69(\pm 0.28)\log(n/3)$, 
with $\epsilon=0.31^{+0.09}_{-0.07}$ dex in $\log M_{\rm bh}$. 

We additionally considered the possibility that the $M_{\rm bh}$--$n$
relation may not be linear, and applied a quadratic fit to account for
the apparent curvature in the distribution.  Although an 
arbitrary empirical function, 
the second order term in the quadratic fit 
\begin{eqnarray}
\log (M_{\rm bh}) & = & 7.98(\pm0.09) +  3.70(\pm 0.46)\log(n/3) \nonumber \\
& & -3.10(\pm 0.84)[\log(n/3)]^2, \nonumber
\end{eqnarray}
with an intrinsic scatter $\epsilon = 0.18^{+0.07}_{-0.06}$ dex, 
is inconsistent with a value of zero at the 99.99\% confidence level.
These relations were constructed with the objective of obtaining 
a relation that can be used to predict SMBH masses in other galaxies
for which the bulge S\'ersic index is known. 
Unlike the linear relation, the quadratic relation predicts finite 
SMBH masses at the high-$n$ end, and $\sim 10^3 M_{\sun}$ mass black 
holes when $n \sim 0.5$, rather than $\sim 10^5 M_{\sun}$ obtained 
with the linear relation. 

The S\'ersic index, a quantity obtained from uncalibrated images, and
independent of galaxy distance, offers an easy way to acquire accurate
estimates of black hole masses in other galaxies.  Moreover, the strength
of the correlation ($r=0.88, r_s=0.95$) implies a fundamental
connection that theory is yet to explain.  In addition, we have shown
that the nuclear star clusters at the centers of low-luminosity bulges
appear to follow the same log-quadratic $M_{\rm bh}$--$n$ relation.
This is suggestive, although not conclusive, that a similar formation
mechanism may be responsible for the build up of massive compact
objects (whether black holes or star clusters) at the centers of
bulges.

Finally, in Section~\ref{Secmu0} we have derived, for the first time,
the expected linear and quadratic relation connecting the mass of a
SMBH with the central surface brightness of its host bulge
(Figure~\ref{Fig-M-mu0}).

%

\acknowledgments

We are grateful to Rebecca Koopmann who kindly supplied us with 
data for NGC~4649 (Koopmann, Kenney, \& Young 2001), 
and to Helmut Jerjen who provided light-profiles for NGC~3115, 
NGC~4486, NGC~4649, and NGC~4697.  The light-profile for 
NGC~4564 was originally obtained by Peter Erwin as a part of
Graham et al.\ (2001).  
We thank Eric Feigelson and Charles Jenkins for 
the statistical advice they provided early on in this work. 
We also thank Stuart Wyithe for his comments and 
kindly supplying us with a code which we subsequently modified 
to fit a quadratic to our data set. 
This research has been supported by the Australian Research
Council through Discovery Project Grant DP0451426.
This research has made use of the NASA/IPAC Extragalactic Database (NED), 
the Isaac Newton Group and HST data archives, 
and StatCodes available at (http://www.astro.psu.edu/statcodes/).

\newpage
\section{APPENDIX A: Galaxy light-profiles}

We have made use of the $R$-band galaxy images of NGC~4486, NGC~4564,
NGC~4649, and NGC~4697 taken by Cheng et al.\ (1997), Frei et al.\
(1996), Kuchinski et al.\ (2000), and HST/ACS archive Prposal ID 10003
(PI C.Sarazin), respectively.  In the case of NGC~3115 we used an
$I$-band image from Kuchinski et al.\ (2000), and for the Milky Way we
used the near-infrared light-profile from Kent, Dame, \& Fazio (1991).

For the $R$-band images, the light-profiles were obtained in the standard way.
Foreground stars and background galaxies were carefully masked, and
the sky-background flux was determined from the mean of $\sim$10
median fluxes that were obtained from small boxes we positioned near
the (galaxy-free) corners of each chip.  The light-profiles were 
extracted 
using the IRAF task ELLIPSE, with the isophotal position angle and
ellipticity free to vary, but the centroid fixed.

An interesting deviation from the above recipe pertains to the analysis
of the edge-on S0 galaxy NGC~3115.  Its light-profile was
obtained using another isophotal fitting routine written in IRAF
(Jerjen, Kalnajs \& Bingelli 2000).  After the foreground and 
background objects, and non-symmetrical features about the nominal
galaxy centre, were removed from
the image, a symmetrical two-dimensional (2D) model was constructed from
the remaining light distribution by allowing the isophotal ellipticity
and position angle to vary, but keeping the luminosity-weighted centre
fixed.  This process was repeated iteratively until the residuals were
minimised.  This left a residual image displaying only the edge-on
disc of NGC~3115.  The one-dimensional surface brightness profile of
the (symmetrical) bulge was then calculated from this 2D model. 

Foreground stars were fitted with a Moffat function that was used to
quantify the point spread function in each image.  However, 
in practice this wasn't important because in avoiding the 
potential presence of nuclear star clusters or partially-depleted
cores, we first removed the inner $\sim$2 seeing disks from the
galaxy light-profiles before fitting S\'ersic's (1963) function. 

\begin{figure*}
\begin{center}
\begin{minipage}{170mm}
\includegraphics[width=170mm]{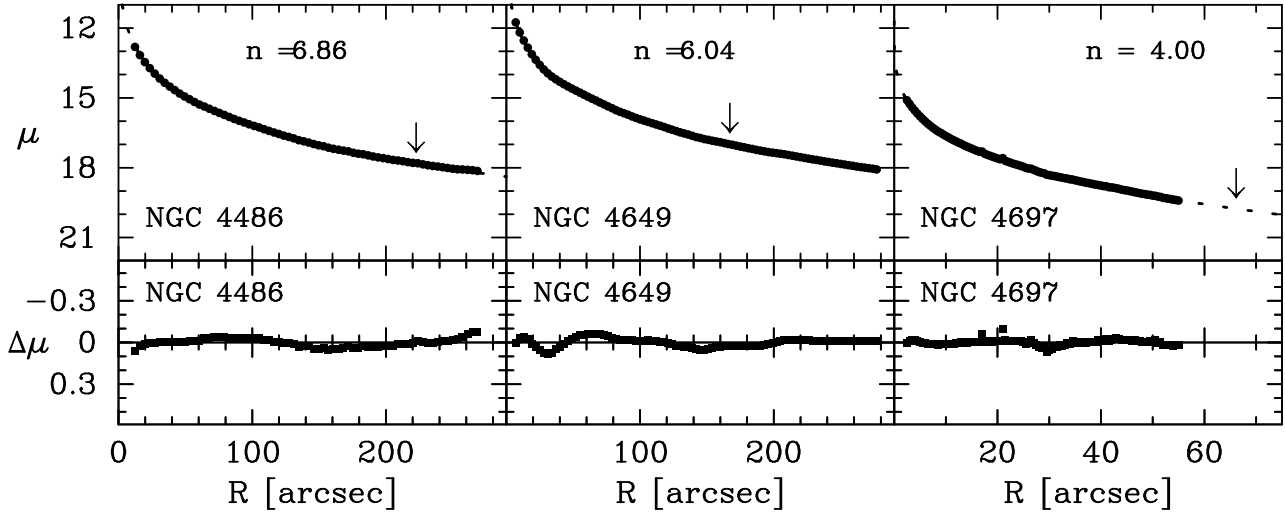}
\vskip30pt
\caption{Major-axis $R$-band light-profiles for three 
elliptical galaxies with directly measured SMBH masses. 
The best-fitting S\'ersic function is shown by the solid curve. 
The arrows mark the spheroids' half-light radii. 
The residuals of the data about the fit are shown in the lower panels.  
(Note: Due to the fact that the $M_{\rm bh}$--$n$ relation does not
require calibrated images, the profiles have not had 
their photometric zero-point determined.)
}
\label{AppFig1}
\end{minipage}
\end{center}
\end{figure*}

\begin{figure*}
\begin{center}
\begin{minipage}{170mm}
\includegraphics[width=170mm]{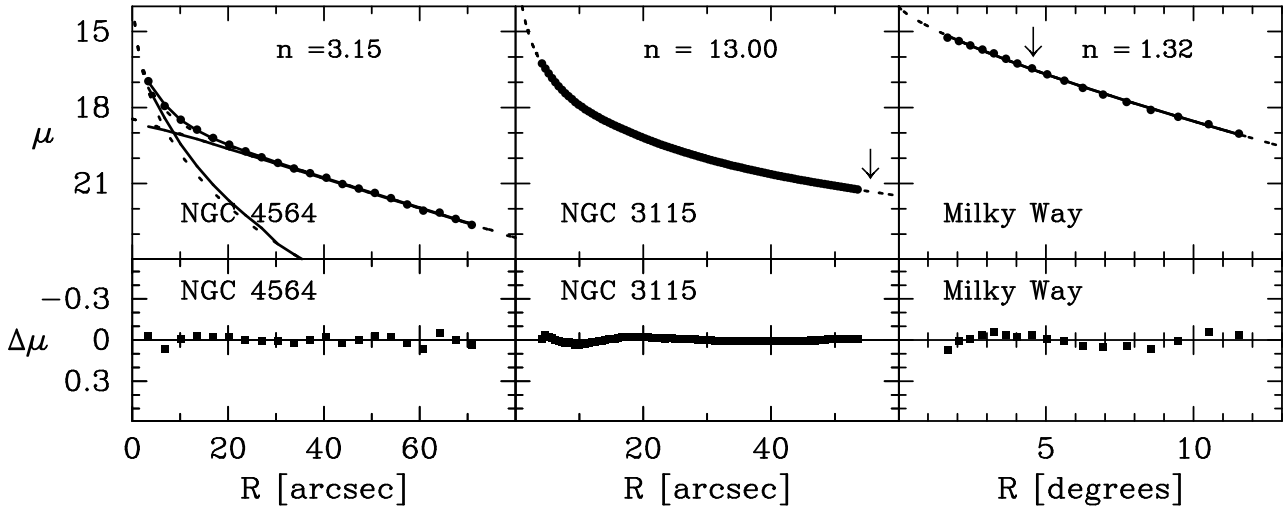}
\vskip30pt
\caption{Major-axis light-profile for the disk galaxy NGC~4564,
the bulge
of the edge-on lenticular galaxy NGC~3115, and the bulge of the 
Milky Way. 
The best-fitting S\'ersic function (and exponential) 
are shown by the curved lines (and the straight lines).
The solid lines show the seeing-convolved functions; the dashed
lines show the unconvolved functions, i.e., the intrinsic profile.  
(Note: Due to the fact that the $M_{\rm bh}$--$n$ relation does not
require calibrated images, the profiles have not 
had their photometric zero-point determined.)
}
\label{AppFig2}
\end{minipage}
\end{center}
\end{figure*}

Each light-profile's best-fitting S\'ersic $R^{1/n}$ function 
plus, in the case of NGC~4564, an exponential disk, was
obtained using the subroutine UNCMND from Kahaner, Moler \& Nash
(1989).  At each iteration, the nonlinear (seeing-convolved) S\'ersic
function (plus exponential function when modelling NGC~4564) 
was approximated by a quadratic function derived from a 
Taylor series.  The quadratic function was minimised to obtain a
search direction, and an approximate minimum of the nonlinear function
along the search direction found using a line search.  The algorithm
computes an approximation to the second derivative matrix of the
nonlinear function using quasi-Newton techniques.
The galaxy light-profiles and best-fitting models are displayed in
Figures~\ref{AppFig1} and \ref{AppFig2}. 
One can see, particularly from the residual profiles in the lower
panels, that the fitted models perform well at matching the 
curvature in the light-profiles.  
This curvature, specifically, the S\'ersic index $n$, is observed to
correlate strongly with a bulge's central supermassive black hole (e.g., 
Figure~\ref{Fig-quad}). 


\label{lastpage}
\end{document}